# Cloud Computing-Future Framework for e-management of NGO's

**Harjit Singh Lamba [1] and Gurdev Singh [2]**

[1] Department Of Computer Science, Eternal University,
Baru Sahib, Himachal Pradesh 173101, India
*harjit.harjit@gmail.com*

[2] Department Of Computer Science, Eternal University,
Baru Sahib, Himachal Pradesh 173101, India
*singh.gndu@gmail.com*

**Abstract**

Cloud computing is an emerging new computing paradigm for delivering computing services. This computing approach relies on a number of existing technologies, e.g., the Internet, virtualization, grid computing, Web services, etc. Cloud Computing aims to provide scalable and inexpensive on-demand computing infrastructures with good quality of service levels. It represents a shift away from computing as a product that is purchased, to computing as a service that is delivered to consumers from the cloud. It helps an organization in saving costs and creating new business opportunities.This paper provides a framework, Education Cloud for the e-management of NGO's. The Education Cloud can transform a nonprofit, or an entire sector of nonprofits, achieves its mission and creates lasting impact in its communities. This paper also presents the case study of Kalgidhar trust, Baru Sahib, Himachal Pradesh, NGO which is using the education as the tool to solve the social issues.

*Keywords:* Cloud Computing, SaaS, NGO, ERP

## 1. Introduction

Cloud Computing is more than a technology. It is more than a platform. It is more than just a hosting provider. It is more than just an application hosted as a service. It is more than providing storage services on the Internet. It is a combination of all the above. A 'cloud' is an elastic execution environment of resources involving multiple stakeholders and providing a metered service at multiple granularities for a specified level of quality (of service).  The definition of cloud computing provided by The National Institute of Standards and Technology (NIST), as it covers, in our opinion, all the essential aspects of cloud computing:
Cloud computing is a model for enabling convenient, on-demand  network access to a shared pool of configurable computing resources( e.g. networks, servers, storage, applications and services) that can be rapidly provisioned and released with minimal management effort or service provider interaction.

Cloud computing employs a service driven business model. Cloud offers services that can be grouped into the following categories
- Infrastructure as a service (IaaS): Hardware resources (such as storage) and computing power (CPU and memory) are offered as services to customers. This enables businesses to rent these resources rather than spending money to buy dedicated servers and networking equipment.. As examples in this category, Amazon1 offers S3 for storage, EC2 for computing power, and SQS for network communication for small businesses and individual consumers.
- Database as a service (DaaS): A more specialized type of storage is offering database capability as a service. Examples of service providers are Amazon SimpleDB, Google BigTable3, Force.com database platform and Microsoft SSDS4. DaaS on the cloud often adopts a multi-tenant architecture, where the data of many users is kept in the same physical table.
- Software as a service (SaaS): In this model, software applications are offered as services on the Internet rather than as software packages to be purchased by individual customers. One of the pioneering providers in this category is Salesforce.com offering its CRM application as a service. Other examples include Google web-based office applications (word processors, spreadsheets, etc.),
- Platform as a service (PaaS): This refers to providing facilities to support the entire application development lifecycle including design, implementation, debugging, testing, deployment, operation and support of rich Web applications and services on the Internet. Most often Internet browsers are used as the development environment. Examples of platforms in this category are Microsoft Azure Services platform6, Google App Engine7, Salesforce.com Internet Application Development platform8 and Bungee Connect platform9. PaaS enables SaaS users to develop add-ons, and also develop standalone Web based applications, reuse other services and develop collaboratively in a team.

Education is the process of training for developing the mind, character and ability to acquire knowledge, analyze it and take necessary steps in time, especially by formal schooling, teaching and training. Thus, the real purpose of education is spread of knowledge as a service to others. A true teacher should place all learning and all efforts sincerely in the service of humanity without the thought of financial gain for self. Education becomes very powerful tool to wipe out the social issues from the society, if implemented with this principle. The NGO's who are based on Education can use the Education Cloud, the cloud based framework as defined in next section.

## 2. Education Cloud

However, cloud computing services could provide many of NGO's with the opportunity to continue to take advantage of new developments in IT technologies at affordable costs. Cloud computing is likely to be an attractive proposition to start-up and small to medium enterprises and educational establishments. The UK's National Computing Center (NCC) estimates that SMEs can reduce the total cost of ownership of technology using hosted solutions [14].
Students in the 21st century have different and vast learning needs which no longer can be satisfied with traditional teaching and learning methodologies [15] i.e. lecture-based, tutorial session, use of multimedia contents and etc. It is essential for Universities and schools to adopt new approach and technology that will better prepare and equip students for current and future job market needs. Education Cloud will address the role of cloud computing technology to improve teaching and learning methodology.

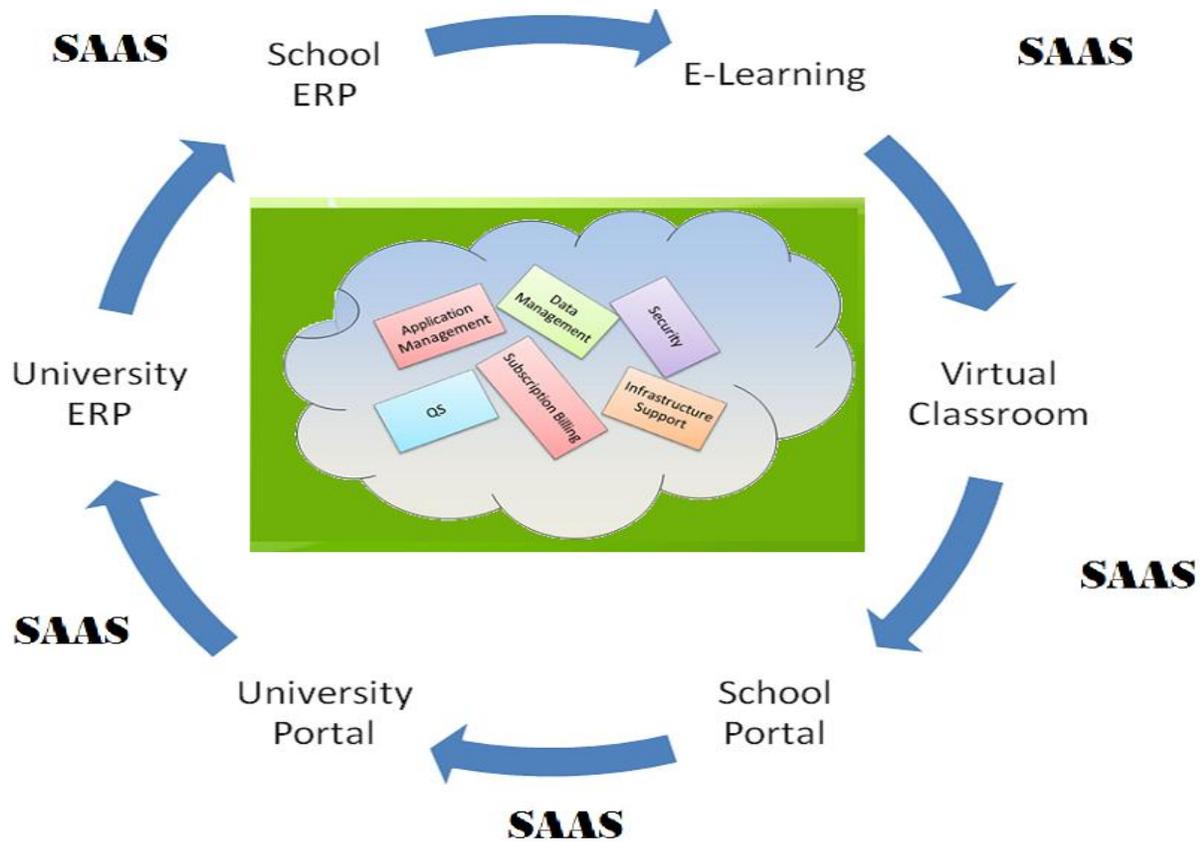

**Figure 1 Education Cloud**

## 3. Advantages of Education Cloud

The advantages of the Education Cloud to the NGO's are studied in the non functional, economic and technological areas.

1. Non-functional Aspects
   - Elasticity – It is an essential core feature of cloud systems and circumscribes the capability of the underlying infrastructure to adapt to changes potentially non functional requirements like size of data supported by an application, number of concurrent users etc. Elasticity does allow the dynamic integration and extraction of physical resources to the infrastructure. Whilst from the application perspective, this is identical to scaling.
   - Reliability – Reliability denotes the capability the capability to ensure constant operation of the system without disruption. It is considered as main features to exploit cloud capabilities.
   - Quality of Service- Quality of service support is a relevant capability that is essential in many cases where specific requirements have to be met by the outsourced services and /or resources. With QoS controls available, cloud providers offer a range of services and price points that provide more choice to customers and back these services with service-level agreements (SLAs) that go beyond uptime and mean time to repair (MTTR) specifications. The result for enterprises is lower-cost IT infrastructure, applicable to a greater range of application types, obtained by combining shared platform economics with high levels of performance assurance.
- Availability – Availability of services and data is an essential capability of cloud systems. It lies in the ability to introduce redundancy for services and data so failures can be masked transparently. With increasing concurrent access, availability is particularly achieved through replication of data /services and distributing them across different resources to achieve load-balancing

2. Economic Aspects
   - Cost Reduction - cloud system can adapt to changing consumer behaviour and reduce cost for infrastructure maintenance and acquisition. Upfront cost to run the system on the cloud is very lower.
   - Pay per use - The capability to build up cost according to the actual consumption of resources is a relevant feature of cloud systems. Pay per use strongly relates to quality of service support, where specific requirements to be met by the system and hence to be paid for can be specified.
   - Improved Time to Market – NGO can focus on meeting their objectives instead of spending time on infrastructure which is not their core competency.
   - Going Green- Clouds principally allow reducing the consumption of unused resources (down-scaling).  Users of cloud computing are more likely to significantly reduce the carbon footprint.

3. Technical Aspects
   - Ease of Use - through hiding the complexity of the infrastructure (including management, configuration etc.)  Cloud can make it easier for the user to develop new applications, as well as reduces the overhead for controlling the system.
   - Location independence: services can be accessed independent of the physical location of the user and the resource.
   - Multi-tenancy - is a highly essential issue in cloud systems, where the location of code and / or data is principally unknown and the same resource may be assigned to multiple users (potentially at the same time).
   - Data Management - As size of data may change at any time, data management addresses both horizontal and vertical aspects of scalability. User need not worry about the database backups.
   - Programming Enhancements – Developers can focus on the business instead of worrying about issues like scalability.

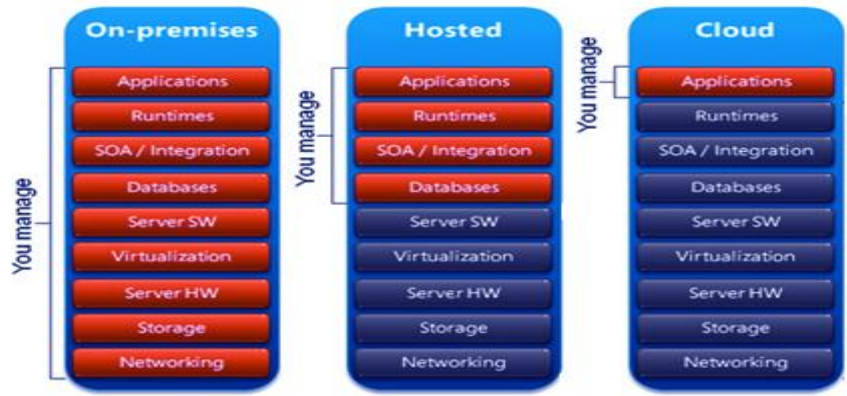

**Figure 1 Cloud Deployment**

### 4. Kalgidhar Case Study

The Kalgidhar Trust is a multifaceted, not for profit charitable organization working for cause of promoting quality education, primary health, social welfare & spiritual uplift to underprivileged/deprived rural masses in far-flung rural areas of North India. Its mission is to establish permanent peace in the world through the synthesis of value based scientific education and moral rejuvenation thereby creating good global citizens.

Kalgidhar Trust started unique revolution in the rural education with their Akal Academies. These academies are affiliated to the Central Board of Secondary Education (CBSE), New Delhi. These academies have a unique academic center, having comprehensive arrangements for imparting worldly scientific education in an environment conductive for spiritual evolution. Currently there are 70 Akal Academies running, spread over northern states of India (Punjab, North Rajasthan, Uttar Pradesh and Himachal Pradesh). All of these academies are located in the backward down-trodden rural areas of North India where education facilities, in general are poor and use of intoxicants is prevalent. For providing the continuity of value based quality education from kindergarten to post-graduate level, Eternal University was set up in April 2008 at Baru Sahib. Eternal University is running 6 colleges with 1500 students. The future plan of the Kalgidhar Trust is to open 500 Akal academies across the rural part of the country and add another 8 colleges to Eternal University in next five years.

**Use of Technology** - Kalgidhar Trust does require a technology solution which can help in running their education centers efficiently. Moreover the solution should be scale with their aggressive growth. Cost of the technology is critical since Kalgidhar Trust works on low cost self-sustaining organization model with less of paid Management, more of dedicated volunteers submitting for whole of life and thus low overheads. The total administration expense of The Kalgidhar Society is only 2.75 % of its annual expenditure.

Educational establishments continue to seek opportunities to rationalize the way they manage their resources. Cloud computing technologies have changed the way applications are developed and used. They are aimed at running applications as services over the internet on a scalable infrastructure. Educations institutions can take advantage of Education Cloud to provide free or low cost platform which will help them to adapt new teaching learning in meeting the student's diverse needs.

### 4.1 Uses of Cloud for Kalgidhar Trust

Seeing the growth of the Kalgidhar trust programs, the scalability provided by Education Cloud is most important benefit to them. By moving from the usual capital upfront investment model to an operational expense, Education Cloud promises to enable Kalgidhar Trust to accelerate the development and adoption of innovative solutions. Kalgidhar Trust can take of advantage of the infrastructure provided by the Education Cloud and focus on building capabilities need to support their objectives. Kalgidhar Trust can use the Education Cloud for the following applications:

1. Integration of all the academies over an Education cloud: - Education Cloud will act as centralized system which connects each academy that make coordination and communication real time. Education cloud will contain all the latest education systems and e learning software. The other academies can use these systems from the centralized system which will reduce the cost of the solution. Moreover, centralized system will have the data of all the academies which will help management in making valuable business decisions in improving the quality of education in all the academies.
2. School management System – ERP system which integrates student, teacher, accounts, library, facilities and administration.

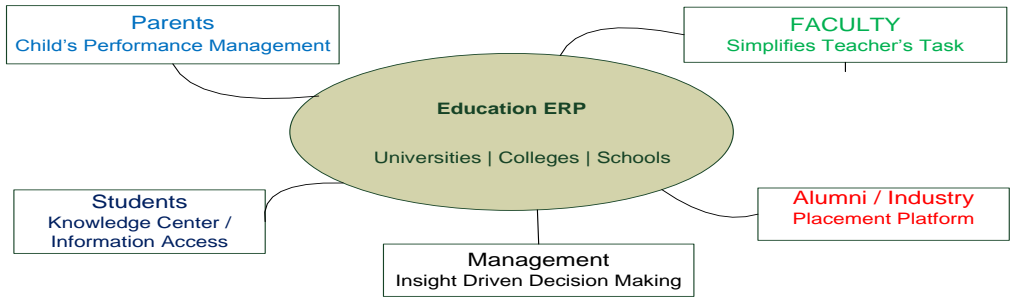

**Figure 2   Education ERP**

3. University management System – ERP like system that integrates and manages admissions, courses, students, teachers, accounts, facilities, research, library etc. The system will automate the admission procedure, registration of students, course details, instructor's details, assignments, grade sheets and transcripts. Students will be able to download their transcripts/grade sheets from online portal.
4. Education Portal- Common place where all the education related resources available from kindergarten to higher studies. This can be used by students, teachers and parents. This can sub divided into school portal and the university portal.
5. E- learning/ Virtual Classroom - online platform to promote communication, collaboration and understanding among students, teachers, subject matter experts and Corporate around the world. This will help the rural areas to get the quality education. It is seen that rural areas have very less quality teachers compared to urban areas. This will reduce this gap. Moreover, the students of the rural areas can learn from the experiences of the experts in the industry.

### 4.2 Business Opportunity

The Kalgidhar Trust is in the education area since 1906. Their strong domain knowledge combine with the Education Cloud can be revenue generator for the Kalgidhar Trust. The education Cloud is formed on the concept of software as service.  The services from Education Cloud can be sold to other schools and universities using a multitenant architecture.  They just need to maintain one more application but can provide lot of value to their customers, i.e. other schools.

### 5.  Conclusion

With help of Education Cloud, the Kalgidhar Trust can focus on the mission of improving lives. Education Cloud can help them in meeting their aggressive future growth of 500 schools. Since all the schools are in the rural areas, this will help in eliminating the social evils from the society. Education Cloud provides cost effective solution in providing the services to support their goal of quality education. Moreover, Education cloud can help them in generating the revenue which can be used on their other philanthropic projects. With the help of the virtual classroom on Education Cloud, the Kalgidhar Trust can start the quality distance education programs which can even reach students of other nations. This will also help to reduce the gap between the university and the corporate world which will eventually help students in meeting their desires of lucrative careers. Other NGO's can use the Education Cloud to support their objectives in spreading the education thru rural world which will help in solving the social evils of the society.